\documentclass{aastex63}

\newcommand{\lamostj}{LAMOST~J~1645+4357}
\newcommand{\hd}{HD~122563}

\received{}
\revised{}
\accepted{}
\submitjournal{ApJ}

\shorttitle{LAMOST J~1645+4357}
\shortauthors{Aoki et al.}

\begin{document}

\title{Unique chemical composition of the very metal-poor star LAMOST J1645+4357
\footnote{}}

\correspondingauthor{Wako Aoki}
\email{aoki.wako@nao.ac.jp}

\author[0000-0002-8975-6829]{Wako Aoki}
\affiliation{National Astronomical Observatory of Japan, National Institutes of Natural Sciences, \\ 2-21-1 Osawa,
  Mitaka, Tokyo 181-8588, Japan
}
\affiliation{Astronomical Science Program, Graduate Institute for Advanced Studies, SOKENDAI, 2-21-1 Osawa, Mitaka, Tokyo 181-8588, Japan
}

\author{Haining Li}
\affiliation{Key Lab of Optical Astronomy, National Astronomical
  Observatories, Chinese Academy of Sciences \\
A20 Datun Road,
  Chaoyang, Beijing 100012, China}

\author[0000-0002-8975-6829]{Nozomu Tominaga}
\affiliation{National Astronomical Observatory of Japan, National Institutes of Natural Sciences, \\ 2-21-1 Osawa,
  Mitaka, Tokyo 181-8588, Japan
}
\affiliation{Astronomical Science Program, Graduate Institute for Advanced Studies, SOKENDAI, 2-21-1 Osawa, Mitaka, Tokyo 181-8588, Japan
}
\affiliation{Department of Physics, Faculty of Science and Engineering, Konan University, 8-9-1 Okamoto, Kobe, Hyogo 658-8501, Japan}

\author{Tadafumi Matsuno}
\affiliation{Kapteyn Astronomical Institute, University of Groningen  \\ Landleven 12, 9747 AD Groningen, The Netherlands}

\author{Satoshi Honda}
\affiliation{Nishi-Harima Astronomical Observatory, Center for Astronomy, University of Hyogo \\ 407-2,
Nishigaichi, Sayo-cho, Sayo, Hyogo 679-5313, Japan}

\author{Gang Zhao}
\affiliation{Key Lab of Optical Astronomy, National Astronomical
  Observatories, Chinese Academy of Sciences \\
A20 Datun Road,
  Chaoyang, Beijing 100012, China}




\begin{abstract}
  We report on the chemical composition of the very metal-poor
  ([Fe/H]$=-2.9$) star LAMOST J1645+4357 that is identified to be a
  red giant having peculiar abundance ratios by Li et al. (2022). The
  standard abundance analysis is carried out for this object and the
  well studied metal-poor star HD~122563 that has similar atmospheric
  parameters. LAMOST J1645+4357 has a remarkable abundance set, highlighted by these features: (1) Nitrogen is significantly enhanced
  ([N/Fe]$=+1.4$) and the total abundance of C and N is also very high 
  ([(C+N)/Fe]=+0.9); (2) $\alpha$-elements are over-abundant with
  respect to iron as generally found in very metal-poor stars; (3) Ti, Sc, Co
  and Zn are significantly deficient; (4) Cr and Mn are enhanced
  compared to most of very metal-poor stars; (5) Sr and Ba are
  deficient and the Sr/Ba ratio ([Sr/Ba]$=-1.0$) is significantly lower than the
  value expected for the r-process. The overall abundance pattern of this object from C to Zn is well reproduced by a faint supernova model assuming spherical explosion, except for the excess of Cr and Mn which requires enhancement of incomplete Si burning or small contributions of a type Ia supernova or a pair-instability supernova. There remains, however, a question why the abundance pattern of this star is so unique among very metal-poor stars. 
\end{abstract}



\section{Introduction} \label{sec:intro}

The chemical abundance ratios of very metal-poor (VMP) stars hold direct evidence of the yields of first generations of massive stars
and supernova explosions \citep[e.g., ][]{McWilliam1995AJ}. Extreme examples are
carbon-enhanced metal-poor (CEMP) stars with [Fe/H]$<4$ \footnote{[A/B]
  = $\log(N_{\rm A}/N_{\rm B}) -\log(N_{\rm A}/N_{\rm B})_{\odot}$,
  and $\log\epsilon_{\rm A} =\log(N_{\rm A}/N_{\rm H})+12$ for
  elements A and B.}  \citep{Christlieb2002Nature, Frebel2005Nature, Keller2014Nature}. The plausible
progenitors of these objects are first generations of massive stars
(10~M$_{\odot} < M < $ 100~M$_{\odot}$) that have ejected little
amount of Fe-peak elements \citep[e.g., ][]{Umeda2003Nature}. Another
interesting example is the very metal-poor star SDSS J0018–0939 that
has low abundances of carbon, $\alpha$-elements and elements with odd atomic numbers, which
might be produced by explosion of a very massive ($M >
$100M$_{\odot}$) star \citep{Aoki2014Sci}. More recently, a clearer chemical signature indicating the existence of very massive stars in the early universe was reported by \citet{Xing2023Nature}.
The latter examples 
particularly have a large impact on the estimates of mass distribution
of first generation stars that are studied by numerical simulations of
star formation from primordial gas clouds.

We conducted high-resolution “snap-shot” spectroscopy with the
Subaru Telescope for candidates VMP stars found by the LAMOST survey.
The observation obtained medium S/N spectra with short exposures
(typically 15-20 minutes) for about 400 stars
\citep{Aoki2022ApJ}. Abundance results for the whole sample were
reported by \citet{Li2022ApJ}. Among the sample studied by the snap-shot
spectroscopy, the extremely metal-poor ([Fe/H]$=-2.9$) red giant star LAMOST J164514.95+435712.0
({\lamostj}) shows unusually low abundances of odd elements (Sc and Co)
and neutron-capture elements (Sr and Ba). These features are similar
to SDSS J0018–0939, although abundances of 
$\alpha$-elements of J1645+4357 are normal. Although carbon is not overabundant, the abundance ratio ([C/Fe]$=0.33$) is higher than those found in typical very metal-poor red giant stars. This star could also be
affected by some unusual supernova explosions.

This paper reports on the abundance results for {\lamostj} from follow-up observations to obtain high-resolution blue spectra. The observations and radial velocities obtained from our spectra and by previous work are reported in \S~\ref{sec:obs}. The abundance analysis and results for individual elements, as well as comparisons with previous work are presented in \S~\ref{sec:anares}. In \S~\ref{sec:disc}, we discuss the distinct features of the abundance ratios of {\lamostj} and constraints on properties of the progenitor by comparing the abundance results with
supernova models.

\section{Observations}\label{sec:obs}

The first high-resolution spectrum of {\lamostj} was obtained with
the Subaru Telescope High Dispersion Spectrograph (HDS;
\cite{Noguchi2002PASJ}) in observing programs for follow-up
spectroscopy of metal-poor star candidates found with LAMOST
\citep{Aoki2022ApJ}. The snap-shot spectrum with $R=36,000$ for
4030--6800~{\AA} was obtained in May 2014 with a short exposure
(15 minutes).  The abundance result obtained from this snap-shot
spectrum was reported by \citet{Li2022ApJ}. To study more detailed
chemical compositions of this star, a high-resolution spectrum of the
blue range (3500--5200~{\AA}) with $R=60,000$ was obtained with
Subaru/HDS on August 30, 2015 with exposure time of 180 minutes. 

Standard data reduction procedures were carried out with the IRAF
echelle package\footnote{IRAF is distributed by the National Optical
  Astronomy Observatories, which is operated by the Association of
  Universities for Research in Astronomy, Inc. under cooperative
  agreement with the National Science Foundation.}. The
signal-to-noise ratios of the spectrum (per 1.8~km~s$^{-1}$ pixel) is 120
and 200 at 4000 and 5000~{\AA}, respectively.

The heliocentric radial velocity is measured from isolated spectral lines used for abundance analysis. The result is given in Table~\ref{tab:rv} with previous measurements with Subaru \citep{Aoki2022ApJ}, with Lick/APF by \citet{Mardini2019ApJ} and those reported in LAMOST DR5. No significant variation of the radial velocity is found between the two measurements with Subaru, nor in all measurements taking the errors in the LAMOST measurements into consideration. This indicates that there is no signature of binarity for this object from radial velocity measurements to date. We note that {\lamostj} is included in the sample of \citet{Price-Whelan2018AJ} who search for binaries from the APOGEE DR14 data, and is not identified as a "high-$K$ star" (a star likely has a companion). 

The atmospheric parameters (effective temperature, surface gravity and
metallicity) of LAMOST J1645+4357 are very similar to those of the
well-known metal-poor star HD~122563 \citep{Wallerstein1963ApJ, Honda2006ApJ}. We select this object as the
comparison star for the abundance analysis in the present work. The spectra of HD~122563
obtained with the Subaru/HDS with $R=$90,000 \citep{Honda2004ApJ} are adopted for this purpose.

\section{Abundance analysis and results}\label{sec:anares}

\subsection{Abundance analysis}\label{sec:ana}

We measure equivalent widths by fitting a Gaussian profile. Atomic
line data for spectral features are taken from previous studies including \citet{Aoki2013AJ} and those taken from VALD
\citep{Kupka2000BaltA}. The measured equivalent widths are given in
Table~\ref{tab:ew}, together with the line data used in the abundance
analysis and sources of the {\it gf} values. We exclude strong absorption lines
($\log(W/{\lambda})>-4.7$) from the analysis because they are not
sensitive to the elemental abundances due to saturation of
absorption. For Na, Al and Si, stronger lines are also used since number of weak
absorption lines useful for abundance analysis is not sufficient. Details on the
spectral lines for individual elements are reported below.

The solar abundances of \citet{Asplund2009ARAA} are adopted to calculate the
[X/Fe] values.

We determine the elemental abundances of LAMOST J1645+4357 and
HD~122563 by 1D/LTE standard analysis and spectrum synthesis
techniques using model atmospheres of the ATLAS NEWODF grid
(\cite{Castelli2003IAUS}).

The effective temperature of LAMOST J1645+4357 (4660~K) is taken from
\citet{Li2022ApJ} who determined it from the color $V-K$. The surface
gravity is derived using the distance information from Gaia EDR3
parallax ($\log g=1.01$) that is provided by \citet{Li2022ApJ}. We
note that \citet{Li2022ApJ} adopted $\log g$ value derived from
spectroscopic analysis  with correction ($\log g=1.18$: see
\citet{Li2022ApJ} for the details) because the Gaia EDR3 data for this
object became available after completing their analysis.  The
metallicity is determined by the Fe abundances from our analysis
([Fe/H]$=-2.86$). This result agrees with the Fe abundance
([Fe/H]$=-2.87$) determined by \citet{Li2022ApJ}.

The atmospheric parameters of {\hd} are adopted from
\citet{Honda2006ApJ}: $T_{\rm eff}=4570$~K, $\log g =1.1$, and
[Fe/H]=$-2.77$. The differences of $T_{\rm eff}$,  $\log g$, and [Fe/H] between {\lamostj} and {\hd} are 90~K, 0.09~dex and 0.09~dex, respectively, demonstrating that {\hd} is suitable for a reference star. \\

\subsection{Abundance results}\label{sec:result}

\begin{figure}
  \includegraphics[width=8cm]{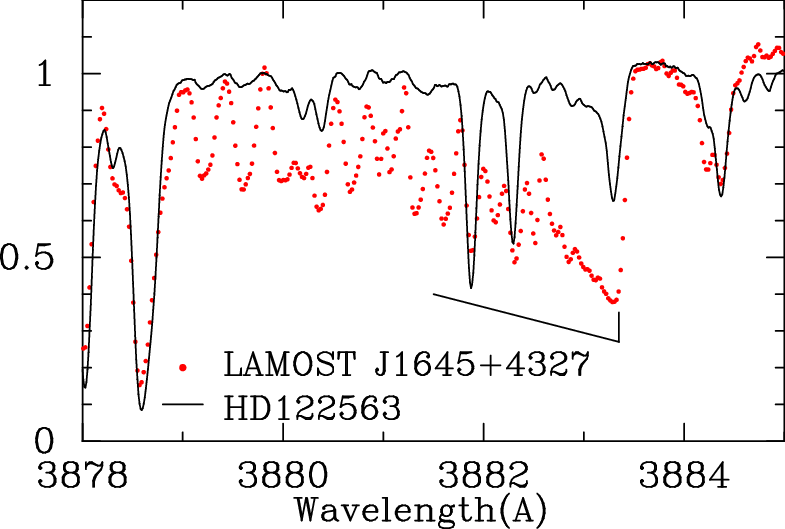}
  \includegraphics[width=8cm]{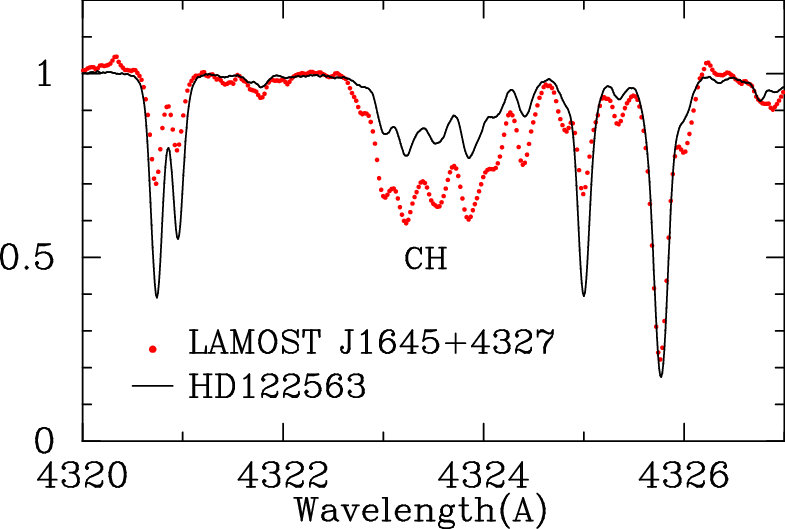}
 \caption{Comparison of spectra of LAMOST J1645+4357 (red filled circles) and HD~122563 (line) for the wavelength ranges including CH and CN molecular bands.}\label{fig:chcn}
\end{figure}

The carbon and nitrogen abundances are determined by the spectrum
synthesis for the CH 4320-4324~{\AA} and CN 3870-3883~{\AA}. The line
data are taken from \citet{Masseron2014AA} and \citet{Brooke2014ApJS}
for CH and CN, respectively. Figure~\ref{fig:chcn} shows comparisons
of spectra between {\lamostj} and {\hd} in the wavelength ranges
covering the CH and CN molecular bands. Both CH and CN bands of
{\lamostj} are clearly stronger than those of {\hd}. The nitrogen
abundance of {\lamostj} is remarkably high ([N/Fe]$=+1.4$). The carbon
abundance of this object, [C/Fe]$=+0.3$, is comparatively high among
highly evolved red giants at this metallicity. The carbon to nitrogen
ratios of the two objects agree well ([C/N]$=-1.1$). This result
suggests that {\lamostj} is originally a carbon-enhanced object, but
the surface material is already affected by the CN cycle and extra
mixing through the evolution in the red giant phase as found in very
metal-poor cool red giants including {\hd} \citep[e.g.,
][]{Spite2005AA}.

The wavelength of the [\ion{O}{1}] 6300~{\AA} is covered by the snap-shot spectrum. The line, however, overlaps with a telluric absorption line and no useful information on the O abundance is derived from the spectrum.

\begin{figure}
  \includegraphics[width=8cm]{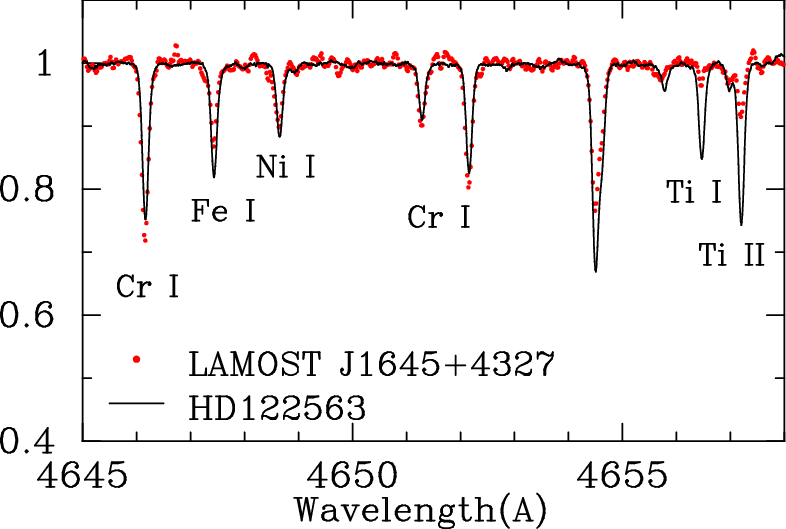}
  \includegraphics[width=8cm]{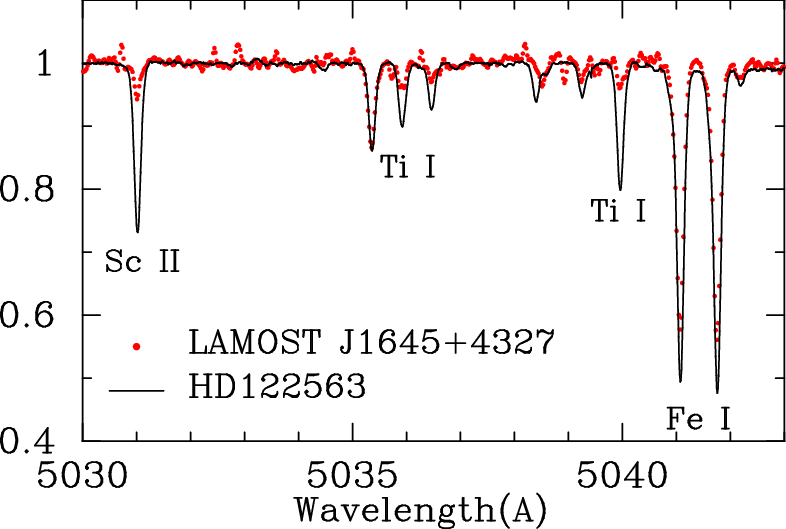}
 \caption{Comparison of spectra of LAMOST J1645+4357 (red filled circles) and HD~122563 (line).}\label{fig:sp2}
\end{figure}

Abundances of $\alpha$ and iron-peak elements are determined by the
standard analysis of measured equivalent widths. The absorption
features of $\alpha$ elements (Mg, Si can Ca) and Fe of the two
objects are very similar, resulting in almost equivalent abundane
ratios between the two stars.  The comparisons of spectra of the two
stars in Figure~\ref{fig:sp2} include Fe lines. Fe lines in {\lamostj}
are slightly weaker than those in {\hd}, resulting in lower
[Fe/H].  We note that slightly higher temperature of {\lamostj} than
that of {\hd} partially contributes to the weaker Fe lines.

The Na abudnances determined from the
\ion{Na}{1} lines also show good agreement. The Al abundances determined from the \ion{Al}{1} 3961 {\AA} are relatively uncertain because the line is severely affected by saturation. The abundance ratios of these two elements ([Na/Fe] and [Al/Fe]) agree within the errors between the two objects.

\ion{Ti}{1}, \ion{Ti}{2} and \ion{Sc}{2} lines of {\lamostj} are
remarkably weaker than those of {\hd} as found in Figure~\ref{fig:sp2}. 
\ion{V}{2} lines detected mostly in the UV range are also weaker in {\lamostj}. 
As a result, the abundance ratios of Sc, Ti and V with respect to Fe of {\lamostj} are significantly lower than those of {\hd}. 

On the other hand, the Cr and Mn abundances of {\lamostj} are higher than those of {\hd}. The excess of Cr absorption lines in {\lamostj} compared to {\hd} is found in Figure~\ref{fig:sp2}. The abundance ratios of Cr and Mn ([Cr/Fe] and [Mn/Fe]) of {\lamostj} are 0.4~dex higher than those of {\hd}. 

We note that, since these two elements are underabundant in {\hd}, the
abundance ratios of {\lamostj} are just slightly higher than the solar
values. Previous studies have reported that Cr abundances derived from \ion{Cr}{1} lines
are systematically lower than those from \ion{Cr}{2} lines for very
metal-poor giants \citep[e.g.,][]{Honda2004ApJ}. The discrepancy is significant in the resonance lines with 0~eV excitation potentail \citep{sneden23}. These lines (4254, 4274, and 4289~{\AA}) are not used in the determination of the final Cr abundance because they are stronger than the criterion in our analysis ($\log(W/\lambda)<-4.7$). The abudances derived from the three 0~eV lines are 0.2~dex lower on average than the final abundance, confirming the result obtained by \citet{sneden23}. Even though these three lines are not includein in our study, the Cr abundances determined
from this species might be underestimated at very low metallicity because of NLTE effect \citep{Bergemann2010AA}. Hence, we here just discuss the relative
abundance of {\lamostj} with respect to {\hd}.

Among the elements heavier than Fe, Co and Zn are significantly
underabundant in {\lamostj} compared to {\hd}, whereas Ni abundances
of the two stars agree very well. The Zn abundance of {\lamostj} is
determined from the \ion{Zn}{1} line at 4722~{\AA}. Another line at
4810~{\AA} that is measured for {\hd} is not available because the
wavelength is affected by the CCD bad column for {\lamostj}. The
difference of the Zn abundances in these two stars is, however,
significant compared to the measurement error.

The Sr abundances are determined from the two resonance lines of
\ion{Sr}{2} at 4078 and 4215~{\AA}. The [Sr/Fe] of {\lamostj} is
remarkably low, whereas that of {\hd} is high among very metal-poor
stars as studied in detail by \citet{Honda2006ApJ}. We note that the
two Sr lines are clearly identified but are not severely saturated in
{\lamostj}, indicating that the derived abundances are reliable.

The Ba abundances are determined from four \ion{Ba}{2} lines for
{\lamostj}. The effect of hyperfine splitting is included in the
calculation using the line data provided by \citet{McWilliam1998AJ}
assuming the Ba isotope ratios of the r-process component of
solar-system material. The effect is only 0.05~dex for {\lamostj}. The
effect is larger (0.26~dex) for {\hd}, because the two resonance lines of this object are stronger than those of {\lamostj}, and one of the two weaker lines used for {\lamostj} is not available for the analysis of {\hd}. Ba is underabundant in this object as well as in {\hd}. 

No other neutron-capture elements than Sr and Ba are detected in the spectrum of {\lamostj}. Upper limits of La and Eu abundances are estimated and given in Table~\ref{tab:abund}.

The error of the abundance of each element is also given in the
table. The random
errors in the measurements ("err" in the table) are estimated to be $\sigma N^{-1/2}$,
where $\sigma$ is the standard deviation of derived abundances from
individual lines, and $N$ is the number of lines used. The $\sigma$ of
Fe {\small I} ($\sigma_{\rm Fe I}$) is adopted in the estimates for
element X for which the number of lines available in the analysis
($N_{\rm X}$) is small (i.e. the error is $\sigma_{\rm Fe I} N_{\rm
  X}^{-1/2}$).  Exceptions are Na and Al abundances, which are
determined from strong lines. We adopt 0.28 and 0.4~dex for the random errors
for these two elements, respectively.

The errors due to the uncertainty of the atmospheric parameters are
estimated for a giant, for $\delta T_{\rm eff}= 100$~K, $\delta \log
g=0.3$, and $\delta v_{\rm turb}=0.3$~km~s$^{-1}$. The total errors are given in Table~\ref{tab:abund}. The error is obtained by adding in quadrature the random error
and errors due to the uncertainties of stellar parameters.

\subsection{Comparison with previous work}\label{sec:prev}

The abundances of 13 elements of {\lamostj} are reported by
\citet{Mardini2019ApJ}. The effective temperature (4810~K) and $\log
g$ (1.39) adopted in their work based on colors and abundance analysis
for \ion{Fe}{1} and \ion{Fe}{2} lines are slightly higher than our
values. The extremely low metallicity ([Fe/H]$=-2.97$), carbon excess
([C/Fe]$=+0.45$), low abundances of Sc and Ti ([Sc/Fe]$=-0.66$, and
[Ti/Fe]$=-0.10$ and $-0.21$ from \ion{Ti}{1} and \ion{Ti}{2}) are
confirmed by the present work. The very low Ba abundance is also found
by \citet{Mardini2019ApJ} ([Ba/Fe]$=-1.01$), although our result is
even lower.

The Cr abundance obtained by \cite{Mardini2019ApJ} ([Cr/Fe]$=-0.43$,
$\log\epsilon=2.24$) is lower by 0.55~dex than our result. Four of the
six \ion{Cr}{1} lines used in their analysis are also measured by our
study, but are not adopted to determine the Cr abundance because they
are stronger than the criterion adopted in the present work. The four lines include the three resonance lines from which lower Cr abundances are derived (see
above). The Cr abundance is determined from nine weaker lines in our analysis. These lines 
are also applied to the analysis of the comparison star {\hd}, reproducing the Cr abundance obtained by previous studies. Hence, the relatively high Cr abundance of {\lamostj}
obtained by the present work is robust.

The low Co abundance determined by the
present work ([Co/Fe]$=-0.58$, $\log\epsilon=1.55$) is not found by
\cite{Mardini2019ApJ} ([Co/Fe]$=0.02$, $\log\epsilon=2.04$). Their
result is obtained from one line of \ion{Co}{1} at 4092~{\AA}. The Co
abundance obtained from this line by the present work is
$\log\epsilon=1.77$, which is the highest among the abundances from the four lines used in
our analysis. Hence, the discrepancy of the Co abundance between the
two studies is partially due to the choice of spectral line. 

\section{Discussion and concluding remarks}\label{sec:disc}

Figure~\ref{fig:ratio} exhibits abundance ratios of six elements for {\lamostj} and {\hd} compared with those of very metal-poor stars studied by \citet{Li2022ApJ}. Similar figures are presented by \citet{Li2022ApJ}, but the values for {\lamostj} are updated by the present work. The comparisons of abundance ratios of these elements demonstrate the uniqueness of {\lamostj} among very metal-poor stars.

\begin{figure}
  \includegraphics[width=12cm]{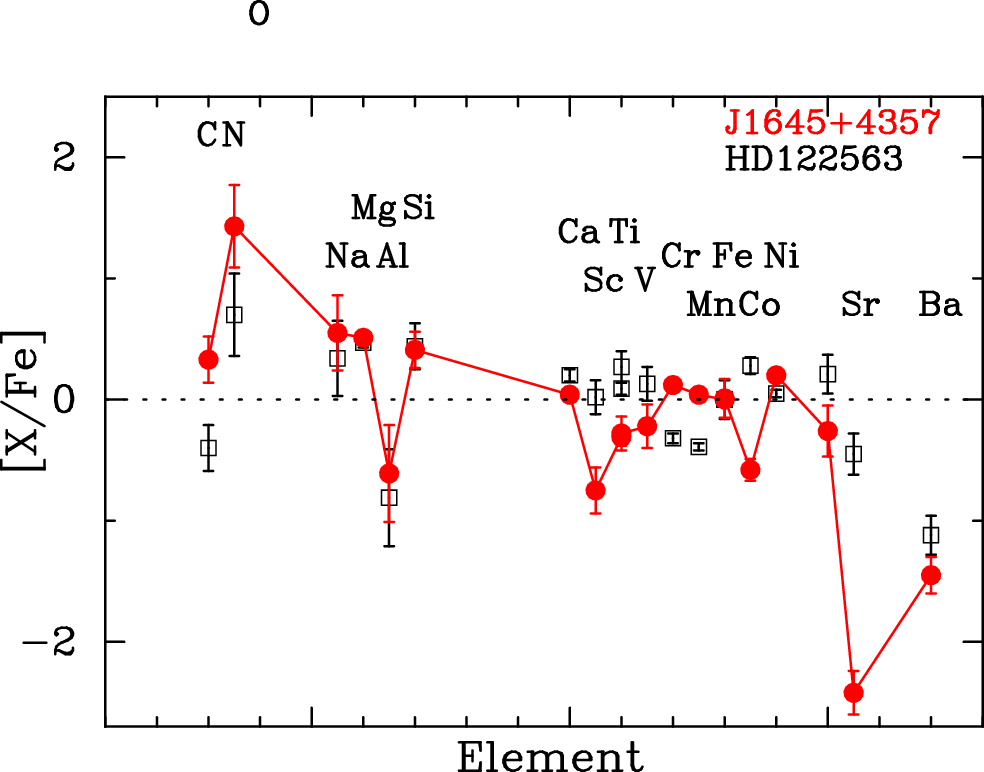}
 \caption{The abundance patterns ([X/Fe]) of LAMOST J1645+4357 (filled circles with line) and HD~122563 (open squares). The error bars indicate the errors of [X/Fe] including the uncertainties of atmospheric parameters.}\label{fig:abundance}
\end{figure}

\begin{figure}
  \includegraphics[width=12cm]{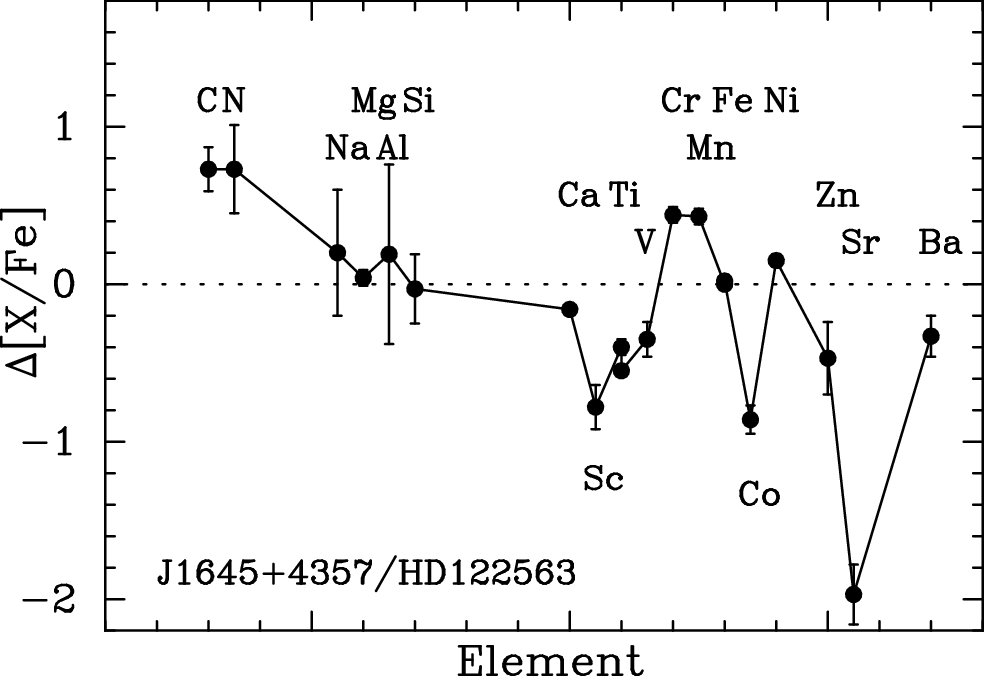}
 \caption{The difference of [X/Fe] values between the two stars ([X/Fe]$_{\rm LAMOST J1645+4357}$ $-$ [X/Fe]$_{\rm HD~122563}$. The error bars indicate the root sum of the quares of random errors estimated for the two objects.}\label{fig:delta_abundance}
\end{figure}

Figure~\ref{fig:abundance} shows the abundance pattern ([X/Fe] as a function of atomic number) for {\lamostj} and {\hd}. Figure~\ref{fig:delta_abundance} shows the difference of [X/Fe] values between the two stars ([X/Fe]$_{\rm LAMOST J1645+4357}$ $-$ [X/Fe]$_{\rm HD~122563}$).

\subsection{Carbon and neutron-capture elements}

The luminosity of {\lamostj} ($\log L=3.3$L$_{\odot}$) indicates that
this object is a highly evolved red giant among extremely metal-poor
stars.  The C abundance ratios of most of such evolved stars,
including {\hd}, are lower than the solar value (i.e., [C/Fe]$<0$) as
demonstrated by \citet{Li2022ApJ}. Adopting the criterion of
CEMP stars including the effect of internal
mixing in red giants, as proposed by \citet{Aoki2007ApJ}, {\lamostj}
is classified into a CEMP star. The large excess of N supports the
interpretation that the C abundance ratio is as high as [C/Fe]$\sim
+1$ in the main-sequence phase of this object. Hence, although this star could be nominally classified into a Nitrogen-Enhanced Metal-Poor (NEMP) star, we regard it as a CEMP star. The no excess of Ba indicates that
this star is a CEMP-no star. CEMP-no stars with moderate excess of
carbon are frequently found in [Fe/H]$<-2.5$ (Placco et al. 2014).

The very large abundance difference of Sr between the two stars is
because of the high Sr abundance of {\hd} among very metal-poor stars.
In addition, {\lamostj} is one of the most Sr deficient objects in
very metal-poor stars ([Sr/Fe]$=-2.4$). The Sr/Ba ratio
([Sr/Ba]$=-1.0$) is lower than the value of the r-process component of
solar-system material ([Sr/Ba]$\sim -0.4$). Whereas stars with high
Sr/Ba ratios are explained by contributions of some process that
provides light neutron-capture elements (e.g., LEPP, weak r-process),
the low Sr/Ba is expected from the s-process (main s-process) at low metallicity. At the very low metallicity, however, contributions of the main s-process by low-mass or intermediate-mass AGB stars are not expected. Hence, the very low Sr/Ba in this object is a problem to be solved by nucleosynthesis or chemical evolution models for the very early Galaxy. We note that the both Sr and Ba abundances of this object are very low, and, hence, a very small contribution of the s-process could significantly change the Sr/Ba ratio. 

Although the origin of neutron-capture elements is unidentified, the
low abundances of Sr and Ba suggest that the chemical composition of
{\lamostj} is very primitive and would be determined by almost a single event.

\subsection{Properties of the progenitor}

\begin{figure}
\includegraphics[width=12cm]{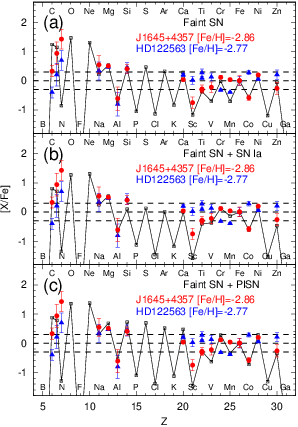}
 \caption{Elemental abundance ratios of {\lamostj} and {\hd} compared with models for (a) a faint SN, (b) a faint SN and a Type Ia SN, and (c) a faint SN and a PISN.}\label{fig:model}
\end{figure}

The high abundance ratios of C and N with respect to Fe in {\lamostj} might be explained by a so-called faint supernova that yields relatively small amount of Fe \citep{Umeda2003Nature}. The models of faint supernovae have been applied to explanations of CEMP-no stars.  Panel (a) of Figure~\ref{fig:model} shows comparisons of supernova yield models with the abundance ratios of {\lamostj} and {\hd}. 
The low abundances of Sc, Ti, V, Co and Zn, as well as the overabundances of $\alpha$-elements, are naturally explained by a quasi-spherical supernova explosion of 25~$M_\odot$ with relatively low explosion energy of $10^{51}$~ergs and a small amount of Fe ejection of $0.0078M_\odot$. Explanations of the high abundance ratios of Sc, Ti, V and Co of {\hd} rather need a high-energy aspherical explosion as proposed for the other EMP stars \citep[e.g., ][]{Umeda2003Nature,Tominaga2007ApJ}. 

The Cr abundance shown in the figure is the result of the LTE analysis. Taking account of the non-LTE effect, the Cr of {\hd} would not be underabundant, and that of {\lamostj} would be overabundant. The overabundance of Cr and Mn in {\lamostj} suggests a large contribution of the incomplete Si burning in the explosion.  This result might imply that there is some mechanism that enhances the incomplete Si burning layer in the projenitor of {\lamostj}. In contrast to the other EMP stars requiring aspherical explosions, the wide incomplete Si burning layer might result from a quasi-spherical explosion with large fallback. Further studies with self-consistent supernova explosion simulations are demanded to investigate the possible connection between incomplete Si burning and sphericity of the explosion. 

An alternative possibility is to assume a contribution of another supernova. Panel (b) of Figure~\ref{fig:model} shows a comparison of the observed abundance pattern with those obtained by a combination of a faint supernova and a type Ia supernova (CCD2, \citealt{Iwamoto1999ApJS}) that provides Cr and Mn as well as Fe. Here, the faint supernova model ejects a small amount of Fe of $0.0078M_\odot$ and the contribution of type Ia supernova to the total Fe mass in {\lamostj} is $68\%$. This combination results in a better fit to the observed abundance pattern. A difficulty of this explanation is that a contribution of a type Ia supernova has longer time-scale than core-collapse supernovae and, hence, is not expected to be effective at this very low metallicity.  
If the timescale of the formation of this object was sufficiently long as a type Ia supernova has contributed, sources of the r-process and the s-process, which might be a merger of binary neutron stars and AGB stars, respectively, should also be effective. 
The deficiency of neutron-capture elements of this object, however, does not support contributions of these sources. Furthermore, the relative number fraction of Type Ia SN is $2\%$ that requires a fine tuning.

Another possible scenario is a combination of contributions of a faint supernova and a pair-instability supernova (PISN) with the He core mass of $100M_\odot$ \citep{Heger2002}, which provides relatively large amount of iron-peak elements (panel (c) of Figure~\ref{fig:model}). Here the faint supernova model ejects a small amount of Fe of $0.0095M_\odot$ and the contribution of PISN to the total Fe mass in {\lamostj} is $56\%$. The relative number fraction of PISN is only $0.2\%$.  The fine tuning is also required to reproduce the abundance pattern. The abundance pattern of elements obtained from this combination, however, does not agree with the observation as in the case of that of a faint supernova and a type Ia supernova, especially the Mn abundance.  
An advantage of this scenario is that a contribution of a PISN is expected in the very early universe contrary to type Ia supernovae \citep{Salvadori2019MNRAS}.

\begin{figure*}
\epsscale{0.5}
\plotone{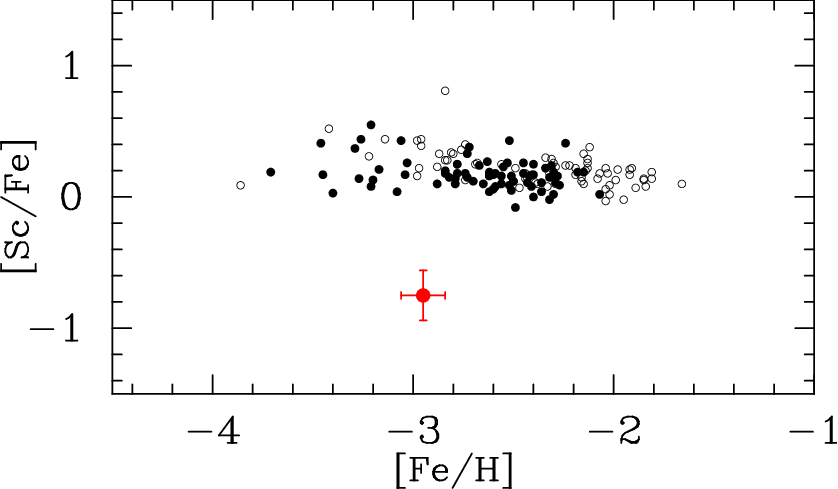}
\plotone{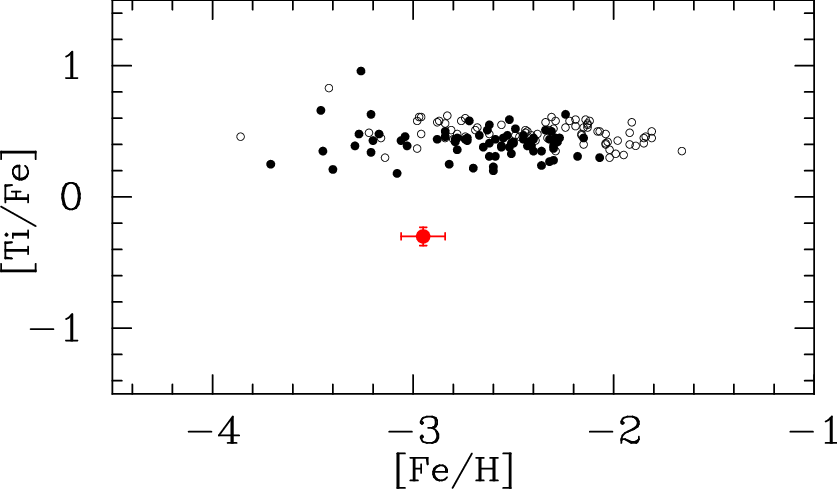}
\plotone{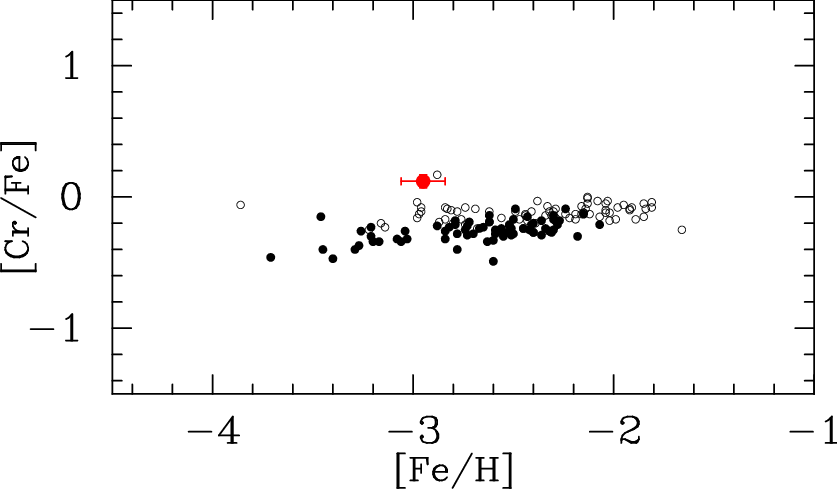}
\plotone{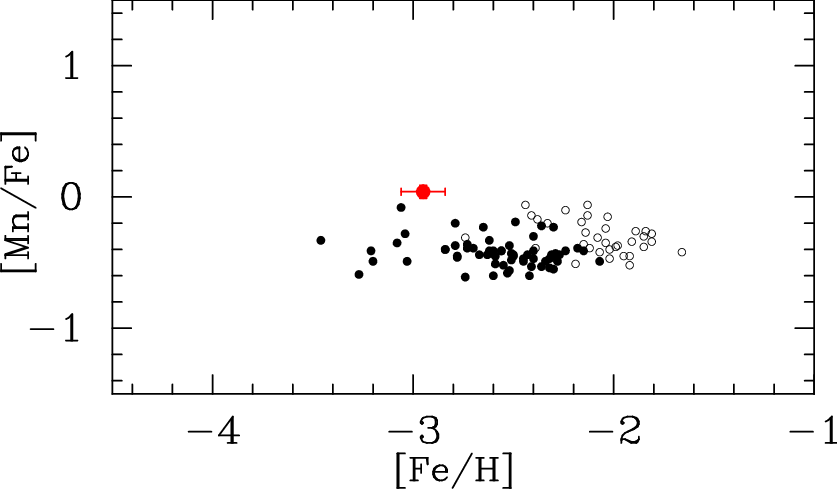}
\plotone{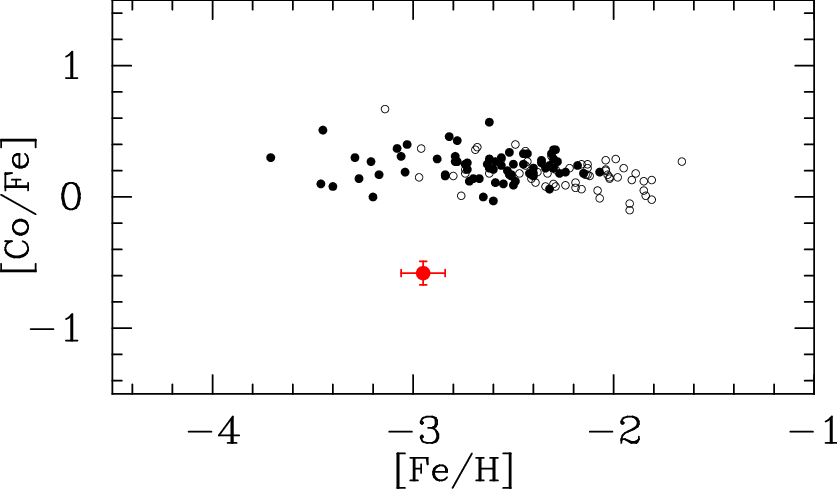}
\plotone{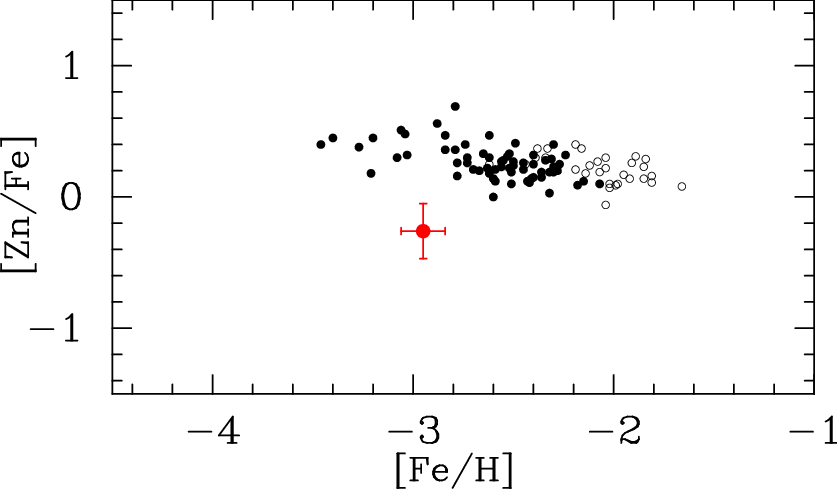}
\caption{Abundance ratios of {\lamostj} (red filled circle with error bars) and very metal-poor stars studied by \citet{Li2022ApJ} (filled circles for stars with $T_{\rm eff}<5500$~K and open circles for stars with $T_{\rm eff}\ge 5500$~K). For the sample of \citet{Li2022ApJ} , stars for which abundances were determined from spectra with signal-to-noise ratios higher than 50 are presented.}\label{fig:ratio}
\end{figure*}

In summary, although further investigation is required for understanding the overabundances of Cr and Mn, the overall abundance pattern of {\lamostj} suggests a contribution of a faint supernova with quasi-spherical explosion. A remaining question is why the abundance pattern of this object is so unique among very metal-poor stars: the abundance distributions of very metal-poor stars presented by \citet{Li2022ApJ} demonstrates the uniqueness of the abundance ratios of {\lamostj} (Fig.~\ref{fig:ratio}). A quasi-spherical explosion of a supernova with low explosion energy is a conventional and natural assumption as a progenitor of very metal-poor stars. The reason why the abundance pattern expected from such an explosion is not observed in other metal-poor stars needs to be explored. The detailed abundance pattern of this star provides a unique opportunity to examine the supernova yields.

\acknowledgments

This research is based on data collected at Subaru Telescope, which is operated by the National Astronomical Observatory of Japan. We are honored and grateful for the opportunity of observing the Universe from Maunakea, which has the cultural, historical and natural significance in Hawaii.
Guoshoujing Telescope (the Large Sky Area Multi-Object Fiber
Spectroscopic Telescope, LAMOST) is a National Major Scientific
Project built by the Chinese Academy of Sciences.  Funding for the
project has been provided by the National Development and Reform
Commission.  LAMOST is operated and managed by the National
Astronomical Observatories, Chinese Academy of Sciences.
This work was supported by JSPS - CAS Joint Research Program. HL and GZ are supported by NSFC No. 11988101. HL is supported by NSFC Nos. 12222305 and 11973049,
and the Youth Innovation Promotion Association of the CAS (id. Y202017). WA, NT and SH are supported by JSPS KAKENHI Grant Numbers 21H04499.

%

\vspace{5mm}
\facilities{LAMOST, the Subaru Telescope}





\bibliographystyle{aasjournal}
\bibliography{J1645ref} 

\clearpage

\clearpage

\begin{deluxetable*}{lrrl}
\tablecaption{Radial velocities\label{tab:rv}}
\tablewidth{0pt}
\tablehead{
\colhead{Obs. date} & \colhead{MJD} & \colhead{$V_{\rm Helio}$} & \colhead{Telescope (ref.)} 
}
\startdata
May  6, 2013 & 56418 & -90.4 & LAMOST (DR5)\\
May 11, 2013 & 56423 & -84.9 & LAMOST (DR5) \\
April 14, 2017 & 57857 & -79.0 & LAMOST (DR5) \\
May 28, 2015 & 57170 & -83.20 & Lick/APF \citep{Mardini2019ApJ} \\
May 10, 2014 &  56787 & -84.4$\pm$0.3 & Subaru \citep{Aoki2022ApJ} \\
August 30, 2015 & 57264  & -84.6$\pm$0.3 & Subaru (This work)\\
\enddata
\end{deluxetable*}

\begin{deluxetable*}{lrrrrl}
\tablecaption{Spectral line data and equivalent widths\label{tab:ew}}
\tablewidth{0pt}
\tablehead{
\colhead{Species} & \colhead{Wavelength (\AA)} & \colhead{$\chi$(eV)} & \colhead{$\log gf$} &\colhead{$W$(m{\AA})} & \colhead{References}
}
\startdata
Na I &  5889.951 &    0.101 &    0.000  &  194.4 & 1 \\
Na I &  5895.924 &   -0.197 &    0.000  &  166.7 & 1 \\
Mg I &  4167.271 &   -0.710 &    4.346  &   48.5 & 2 \\
Mg I &  4571.096 &   -5.688 &    0.000  &   75.0 & 1 \\
\enddata
\tablenotetext{}{References --
(1)\citet{Morton1991ApJS}; 
(2)\citet{Fischer1975CaJPh}; 
(3)\citet{Chang1990JQSRT}; 
(4)\citet{SLa}; 
(5)\citet{BL}; 
(6)\citet{GARZ}; 
(7)\citet{Wiese80}; 
(8)\citet{SN}; 
(9)\citet{SR}; 
(10)\citet{LGWSC}; 
(11)\citet{Blackwell1982aMNRAS}; 
(12)\citet{Blackwell1982bMNRAS}; 
(13)\citet{SLS}; 
(14)\citet{MFW88}; 
(15)\citet{DLSSC}; 
(16)\citet{BWL}; 
(17)\citet{FMW}; 
(18)\citet{BK}; 
(19)\citet{BKK}; 
(20)\citet{Lawler2015ApJS}; 
(21)\citet{WLSCow}; 
(22)\citet{Biemont1980AA}; 
(23)\citet{Lawler1989JOSAB}; 
(24)\citet{WLSC}; 
(25)\citet{BHN}; 
(26)\citet{Ryabchikova1994MNRAS}; 
(27)\citet{Pickering2001ApJS}; 
(28)\citet{Wood2014ApJS};  
(29)\citet{Melendez2009AA};
(30)\citet{Pinnington1995JPhB}; 
(31)\citet{McWilliam1998AJ}; 
(32)\citet{Davidson1992AA}.
(This table is available in its entirety in machine-readable form.)}
\end{deluxetable*}

\begin{deluxetable*}{lccccccccccccccc}
\tablecaption{Abundance results\label{tab:abund}}
\tablewidth{0pt}
\tablehead{
  \colhead{Element} & \colhead{Sun} & \multicolumn{5}{c}{HD122563} & &
  \multicolumn{5}{c}{J1645+4357}  & & \multicolumn{2}{c}{$\Delta$J1645/HD122563}  \\
  \cline{3-7} \cline{9-13} \cline{15-16}
    &  $\log\epsilon$(X) &  $\log\epsilon$(X) & [X/Fe] & $N$ & err & err$_{\rm tot}^{a}$ & & $\log\epsilon$(X) & [X/Fe] & $N$ & err & err$_{\rm tot}^{a}$ && $\Delta_{\rm J1645/HD122563}$ & err$_{\Delta}^{b}$  
}
\startdata
C & 8.43 & 5.30 & -0.40 &  & 0.10 & 0.19 && 5.90 & 0.33 &  & 0.10 & 0.19 && 0.73 & 0.14 \\
N  & 7.83 & 5.80 & 0.70 &  & 0.20 & 0.34 && 6.40 & 1.43 &  & 0.20 & 0.34 && 0.73 & 0.28 \\
Na & 6.24 & 3.85 & 0.34 & 2 & 0.28 & 0.31 && 3.93 & 0.55 & 2 & 0.28 & 0.31 && 0.20 & 0.40 \\
Mg & 7.60 & 5.34 & 0.47 & 4 & 0.02 & 0.04 && 5.25 & 0.51 & 4 & 0.04 & 0.05 && 0.04 & 0.05 \\
Al & 6.45 & 2.91 & -0.81 & 1 & 0.40 & 0.40 && 2.98 & -0.61 & 1 & 0.40 & 0.40 && 0.19 & 0.57 \\
Si & 7.51 & 5.22 & 0.44 & 1 & 0.18 & 0.19 && 5.07 & 0.41 & 2 & 0.14 & 0.15 && -0.03 & 0.22 \\
Ca & 6.34 & 3.81 & 0.20 & 12 & 0.02 & 0.05 && 3.52 & 0.04 & 13 & 0.03 & 0.05 && -0.16 & 0.03 \\
Sc & 3.15 & 0.44 & 0.02 & 4 & 0.04 & 0.14 && -0.46 & -0.75 & 2 & 0.14 & 0.19 && -0.78 & 0.14 \\
Ti I & 4.95 & 2.31 & 0.09 & 30 & 0.02 & 0.05 && 1.78 & -0.31 & 17 & 0.05 & 0.07 && -0.40 & 0.05 \\
Ti II & 4.95 & 2.49 & 0.27 & 29 & 0.02 & 0.13 && 1.82 & -0.28 & 21 & 0.02 & 0.14 && -0.55 & 0.03 \\
V & 3.93 & 1.32 & 0.13 & 5 & 0.03 & 0.14 && 0.85 & -0.22 & 3 & 0.11 & 0.18 && -0.35 & 0.11 \\
Cr & 5.64 & 2.59 & -0.32 & 9 & 0.02 & 0.04 && 2.90 & 0.12 & 9 & 0.05 & 0.05 && 0.44 & 0.05 \\
Mn & 5.43 & 2.30 & -0.39 & 5 & 0.02 & 0.03 && 2.61 & 0.04 & 4 & 0.04 & 0.05 && 0.43 & 0.05 \\
Fe I & 7.50 & 4.77 & 0.00 & 137 & 0.02 & 0.02 && 4.64 & 0.00 & 144 & 0.02 & 0.02 && 0.00 & 0.03 \\
Fe II & 7.50 & 4.76 & 0.00 & 12 & 0.02 & 0.16 && 4.66 & 0.01 & 12 & 0.03 & 0.16 && 0.02 & 0.04 \\
Co & 4.99 & 2.54 & 0.28 & 7 & 0.05 & 0.07 && 1.55 & -0.58 & 4 & 0.07 & 0.09 && -0.86 & 0.09 \\
Ni & 6.22 & 3.54 & 0.05 & 6 & 0.03 & 0.03 && 3.56 & 0.20 & 4 & 0.02 & 0.02 && 0.15 & 0.04 \\
Zn & 4.56 & 2.04 & 0.21 & 2 & 0.13 & 0.16 && 1.44 & -0.26 & 1 & 0.19 & 0.21 && -0.47 & 0.23 \\
Sr & 2.87 & -0.32 & -0.45 & 2 & 0.13 & 0.17 && -2.41 & -2.42 & 2 & 0.14 & 0.18 && -1.97 & 0.19 \\
Ba & 2.18 & -1.67 & -1.12 & 3 & 0.10 & 0.16 && -2.13 & -1.45 & 4 & 0.08 & 0.15 && -0.33 & 0.13 \\
La & 1.10 & -2.50 & -0.87 &   &  & & & $<-2.50$ & $<-0.74$ &  &  &  &&  &  \\
Eu & 0.52 & -2.85 & -0.64 & 3 &  & & & $<-3.20$ & $<-0.86$ &  &  &  &&  &  
\enddata
\tablenotetext{a}{Errors of [X/Fe] values including random errors and errors due to uncertainties of atmospheric parameters.}
\tablenotetext{b}{Errors obtained by adding in quadrature the random errors of the two stars.}
\end{deluxetable*}








\end{document}